\documentclass[a4paper,12pt]{JHEP3}
\usepackage{amssymb}
\usepackage{mathrsfs}
\usepackage{cite}
\usepackage{texdraw}
\usepackage[english]{babel}

\csname @addtoreset\endcsname{equation}{section}

\newcommand{\bea}{\begin{eqnarray}}
\newcommand{\eea}{\end{eqnarray}}

\title{Revisiting the D1/D5 System \\ or Bubbling in AdS$_3$}
\author{Matteo Boni\\
Dipartimento di Fisica dell'Universit\`a di Milano and\\
INFN, Sezione di Milano,\\
Via Celoria 16, I-20133 Milano, Italy.\\
E-mail: \email{matteo.boni@mi.infn.it}}

\author{Pedro J. Silva\\
INFN, Sezione di Milano, \\ 
Via Celoria 16, I-20133 Milano, Italy and\\
Institut de Fisica d'Altes Energies (IFAE),\\
Edf. Cn, Universitat Autonoma de Barcelona (UAB),\\
E-08193 Bellaterra (Barcelona), Spain.\\
E-mail: \email{pedro.silva@mi.infn.it}}

\preprint{IFUM-835-FT\\hep-th/0506085}

\abstract{In this article we study the relation between the
bubbling construction and the Mathur's microscopic solutions 
for the D1/D5 system.
We have found that the regular near horizon D1/D5 system (after
appropriated constraints are imposed) contains all the bubbling
regular solutions. Then, we show that the features of this system
are rather different from the bubbling in $AdS_5\times S^5$, since the perimeter
and not the area plays a key role. After setting the main
dictionary between the two approaches, we investigate on extensions
to non-regular solutions like conical defects and/or naked
singular solutions. In particular, among the latter metrics, closed
time-like curves are found together with a chronology protection
mechanism enforced by the AdS/CFT duality.}


\begin{document}

\section{Introduction}

String theory is the most promising candidate to accommodate all
the fundamental forces of the universe, including gravity.
Unfortunately, our understanding of this theory is far from
complete. In particular, how quantum gravity and the standard
model are realized within string theory is still elusive.

However, we do not need to solve in full string theory to
learn important lessons: in some cases, even simple toy models
produce amazing physical outputs. This is the case of the AdS/CFT
conjecture, where we can study closed string theory (and therefore
quantum gravity) as dual to super Yang-Mills theory (SYM).
Actually what we really mean is that within this framework, we
can concentrate on a sector of the theory where a lot of control
is achieved but physical relevance is still present.

Among the many different possibilities, our studies are motivated
by recent works on 1/2 BPS sectors of near horizon geometries
sourced by a stack of $N$ D3-branes \cite{Corley:2001zk,Berenstein:2004kk,
Caldarelli:2004ig,Lin:2004nb}. In this case, the
{\em bubbling construction} consists in looking for solutions of type
IIB supergravity theory, where only the metric and the self dual
Ramond-Ramond 5-form are excited. Furthermore, it is required to have
regular solutions with an $SO(4) \times SO(4)$ symmetry and at least
1/2 BPS supersymmetry. Once the above solutions are found, by
means of the AdS/CFT duality, the corresponding dual operators in
$N$=4 SYM theory are identified. One of the most interesting
outcomes of the above studies is the simplicity of the dual field
theory, a matrix quantum mechanics, that gives the possibility to
obtain a deeper understanding on fundamental issues of quantum
gravity: for example the role of closed time-like curves
(CTC) \cite{Caldarelli:2004mz}, the study of black hole thermodynamics 
\cite{Gubser:2004xx} and the appearance of stretched horizons 
\cite{Suryanarayana:2004ig} probing black hole entropy laws, 
and even the structure of the quantum phase space \cite{Mandal:2005wv ,Grant:2005qc}.

Although there is a growing amount of literature on this subject,
we noticed that the other superconformal cases, M2, M5, D1/D5, are
less understood and therefore they deserve some attention. Following
the above reasoning, we study the AdS/CFT conjecture for the D1/D5
case, focusing on the simplest 1/4 BPS sector (from the ten dimensional point
of view) with an $SO(2)\times SO(2)$ symmetry, that is known as
{\em bubbling in $AdS_3$}.

Bubbling in $AdS_3$ has already been studied, to the best of our
knowledge, in three papers \cite{Liu:2004ru,Martelli:2004xq,Liu:2004hy}. 
First we have two papers
produced by Liu et al, that consider from first
principles mainly the supergravity side, leaving open the relation
with the CFT theory. Secondly, the work of Martelli and Morales, 
that uses an already known family of solutions to obtain
the supergravity equations. Then, they make a conjecture relating
the regular bubbling solutions to the known regular solutions of 
the D1/D5 system constructed by the group of Mathur 
\cite{Lunin:2002bj,Lunin:2002fw, Lunin:2001jy}. We believe that there
are still many important points that need to be clarified. In particular,
before deeper studies can be done, we need to set the basics of
the AdS/CFT dictionary and to get a better understanding on the
relation between the bubbling supergravity solutions and the
related dual CFT operators, including a study of regularity 
conditions and of the way other non regular geometries could
appear in this framework. The scope of the present work is
precisely to add on this direction.

The paper is organized as follows: in section \ref{sugra} we
review the form of the bubbling ansatz from \cite{Liu:2004hy}, paying
particular attention to the field equations and their general
solution in terms of two different types of sources or boundary
conditions on a two dimensional plane. Then, adding the regularity
requirement, we are able to reduce the two types of sources into a
single one, constrained to live on a monodimensional curve. We
compute the total flux of the solutions finding that it comes in
terms of a line integral on the source and it is proportional to its
length. In section \ref{d1d5} we review the D1/D5 supergravity
solution, together with the basic input coming from the CFT
theory. Then, we show that it is straightforward to constrain the
D1/D5 solutions to include the solutions discussed in the previous
section, realizing the conjectured relation between the two
approaches presented in \cite{Martelli:2004xq}. At this point, the basics
of the AdS/CFT dictionary for the bubbling construction are given.
In section \ref{cft} we show how non-regular solutions can be
included into this duality study by relaxing the regularity
conditions, in order to enlarge the family of solutions, obtaining conical
singularities, Aichelburgh-Sexl and naked singularities. Some of
these metrics have a well defined dual operator, and some don't.
In particular, we have found solutions with CTC that nevertheless
seem to have no counterpart in the D1/D5 system, realizing a sort
of chronology protection mechanism enforced by string theory. In
section \ref{last} we comment on the CFT dual description and
the possible role of the Liouville theory in the duality, together
with a short discussion on future work.

It is important to highlight that along this article we will be always
working within the minimal supergravity framework. Hence, we only excite
the metric $g$ and the 3-form field strength $H_{(3)}$, to avoid
further complications. Nevertheless, inclusion of tensor multiplets
should not be much more difficult and is left for future works. We
also point out that, although this last reduction seems to exclude
giant graviton configurations (since they have been identified with 
supergravity solutions including non-trivial dilaton field \cite{Lunin:2002bj}), 
we still find states that look very much like a giant graviton, 
at least from the bubbling point of view. 
More on this can be found in section (\ref{last}).


\section{Bubbling ansatz for AdS$_3$}
\label{sugra}

We start this section by reviewing some known facts about bubbling in
six dimensional models (see \cite{Martelli:2004xq,Liu:2004hy} 
for a derivation of the supergravity ansatz). 
The working hypothesis is to look for the simplest states in the 
D1/D5 system, identifying their quantum numbers and symmetries to translate 
them into isometries on the supergravity side, giving form to the 
corresponding ansatz.

In short, for the near horizon D1/D5 system, the spectrum of
chiral primaries is classified by the conformal dimensions
$(h,\bar h)$ and the R-charges $(j, \bar j)$, related to the
symmetries $SO(2,2)\times SO(4)\, \sim SL(2,R)_L\times
SL(2,R)_R\times SU(2)_L\times SU(2)_R$. Here, we are interested in
$N =(1,0)$ minimal six dimensional supergravity, that is universal
for Kaluza-Klein reductions from ten dimensions, either on $T^4$
or on $K3$. The simplest family of states in this setting is given by
$h=\bar h$ and $j=\bar j$
(see \cite{Maldacena:1998bw,deBoer:1998ip,Deger:1998nm} for 
studies on chiral primaries).

The corresponding supergravity ansatz is defined by the above
symmetries and the nature of the chiral states under study. The 
general form of the solution is worked out from first principles in
\cite{Liu:2004ru,Liu:2004hy} or deduced from previously known results, 
in \cite{Martelli:2004xq}. In both
cases the metric is given by\footnote{In this work we will always write 
the six dimensional metric, and not the corresponding ten dimensional 
metric, in order to include the possibility to 
work with either $T^4$ or $K3$.}
\begin{eqnarray}
\label{bubbling1} 
&&ds_6^2=-h^{-2}\left(dt + V\right)^2 + h^2
\left(dy^2+\delta_{ij}\,dx_i dx^i \right)
+ y\left(e^G d\theta_1^2 + e^{-G} \left( d\theta_2 + \chi d\theta_1 \right)^2 \right)\,,
 \nonumber \\
&&e^{-G} \ = \ h^2 y + (h^2 y)^{-1} \left( z - \frac{1}{2} \right)^2 \,,
\qquad \chi \ = -\ e^G \left[ h^2 y + (h^2 y)^{-1} \left( z^2 -
\frac{1}{4} \right) \right]\,, \cr
&&dV=-{1\over y}*_3dz\,.
\end{eqnarray}
where $i=1,2$ and $*_3$ is the three dimensional Hodge dual of flat metric in the space 
directions $(y,x^1,x^2)$. 
The self-dual 3-form field strength $H_{(3)}$ is written in terms of
a 2-form ${\tilde F}_{(2)}$ and the four dimensional Hodge dual $*_4$ of the metric
in the $(t,y,x^1,x^2)$ directions, as follows:
\bea
&&H_{(3)}=-{1\over2}[ F_{(2)}\wedge d\theta_1+{\tilde F}_{(2)}\wedge (d\theta_2+\chi d\theta_1)]\,,
\quad F_2=e^G*_4{\tilde F}_2\,, \cr \cr
&&\tilde{F}_{(2)}  =  -2 \left[ d \left( \left( 1 + \chi \right) y
e^{-G} \right) \wedge \left( dt + V \right) - h^2 e^{-G} *_3
\left( d \left( y e^G \right) + \left( 1 + \chi \right) y e^{-G}
d\chi \right) \right]\,. \nonumber
\eea
At last, we have that $h^2$ and $z$ are constrained by the equations
$$
d[*_3 y d(h^2)] \ = \ 0 \qquad d[*_3 \frac{1}{y} dz] \ = \ 0\,.
$$
Therefore, the whole solution is defined by these two
independent functions $h^2$ and $z$, obeying second order differential
equations.
Expanding these last two equations, we observe that they can be
understood as Laplace equations in four and six dimensional
auxiliary spaces,
\bea
y\,\Delta_{(4)}(h^2) = 0 \quad ,\quad
y\,\Delta_{(6)}\left({z\over y^2}\right)=0\,. \nonumber
\eea
Hence, the general solution can be written in terms of the Green 
functions with sources at the plane $y=0$,
\begin{eqnarray}
h^2(x_1,x_2,y)  =  \int_{{\mathbb R}^2} \frac{\rho(x_1^\prime,
x_2^\prime) dx_1^\prime dx_2^\prime}{(x-x^\prime)^2+y^2} \quad ,
\quad z(x_1,x_2,y)  =  \frac{y^2}{\pi} \int_{{\mathbb R}^2}
\frac{z(x_1^\prime, x_2^\prime) dx_1^\prime
dx_2^\prime}{[(x-x^\prime)^2+y^2]^2}\,. \nonumber 
\end{eqnarray}
Notice that we have introduced $\rho(x^1,x^2)$ as the source only for
$h^2$, since for $z$ the source coincides with its value
at $y=0$ i.e. $z(0,x^1,x^2)\equiv z(x^1,x^2)$.

Up to this point, the above solutions solve the bubbling ansatz part
related to the symmetries. It is still needed to
impose regularity conditions to finish the work. In previous
papers, it was conjectured that such a condition would connect
these solutions to the known solutions of D1/D5 system
characterized by the profile ${\bf F}$ of the corresponding winding
string. This is a natural conjecture since we are just describing the 
same system from a different perspective and, therefore, the two approaches 
have to be connected. In what follows we provide such analysis.

\vspace{1cm}\noindent \textsc{\bf Regularity condition}
\vspace{.5cm}\\
Following the analysis done in the LLM paper \cite{Lin:2004nb}, 
we study how the regularity condition imposes constraints upon 
our metrics. Due to the form of the
solution, the possible conflicting regions are at the $y=0$ plane.
In fact, we can see that in order to have smooth geometries as we
approach the source plane, one or both radii of the circles associated to 
$\theta_1$ and $\theta_2$,
$$
R_1^2 \ = \ y ( e^G + e^{-G} \chi^2 ) \quad \mbox{ and } \quad R_2^2 \ = \ y e^{-G} ,
$$
should mix with the $y$ coordinate and, also, $h^2$ or its inverse has
to be regular. 

We first consider the case where $R_1$ remains constant and
$R_2$ recombines to $y$. In this case, after expanding all the
relevant fields in $y$, we arrive to the following result
$$
\begin{array}{lll}
&&z  =  \frac{1}{2} + z_2 y^2 + O(y^3) \\ \\
&&h^2  =  h_0^2 + h_1^2 y + O(y^2) \\ \\
&&e^{-G}  =  h_0^2 y + h_1^2 y^2 + O(y^3) \\ \\
&&\chi  =  - \frac{1}{h_0^2} \left( h_0^2 + \frac{z_2}{h_0^2} \right) 
+ O(y)\quad\quad\quad\quad
\end{array}
$$
where $f_n={d^n\over dy^n} f|_{y=0}$ and $f$ is any of the
involved functions. The other possibility where we have a
vanishing $R_1$ and a constant $R_2$ gives
$$
\begin{array}{lll}
&&z = - \frac{1}{2} + z_2 y^2 + O(y^3) \\ \\
&&h^2 = h_0^2 + h_1^2 y + O(y^2) \\ \\
&&e^{-G}= \frac{1}{h_0^2 y} -
\frac{h_1^2}{h_0^4} + \left( h_0^2 - 2 \frac{z_2}{h_0^2} \right) + O(y^2) \\ \\
&&\chi  = - h_0^2 \left( h_0^2 + \frac{z_2}{h_o^2} \right) y^2 + O(y^3)
\end{array}
$$
Notice that the above set of equations tells us that $z$ is a
constant on the $y=0$ plane, with values $(1/2,-1/2)$ only. Also,
notice that $\chi$ has a rather different behaviour on the two
regions of the plane, and that up to now, $h^2$ is unconstrained.
We can use $z$ to define two different regions on the plane,
region (I) where $z=1/2$ and region (II) where $z=-1/2$. Since we
are in a two dimensional hypersurface, the frontier has to be a
curve ${\bf C}$.

To complete the regularity analysis, we have to probe the vicinity
of the boundary of the two regions (I) and (II) or, if you prefer,
the neighbourhood of the curve ${\bf C}$. Basically, we need both 
radii to smoothly
combine with $y$ at this locus (as occurs in the bubbling
solutions for the D3-brane case, that results in the pp-wave
solution).

Given a general curve ${\bf C}$, we use that whatever shape it assumes,
we can expand in term of the exterior curvature of ${\bf C}$ and that
locally we can always find adapted coordinates where $x^1$ is
perpendicular to the boundary and $x^2$ is parallel. Then, we
change coordinates to following pp-wave alike coordinates
\bea
\label{pp-wave}
y=r_1r_2\quad , \quad x_2={1\over 2}(r^2_1-r^2_2)
\eea
and take the limit $y\rightarrow 0$ asking for a resulting smooth
geometry. After some algebra, it is not difficult to see that in
order to achieve regularity,
$$
h^2={1\over 2x^2}(1+ O(y^2/x^2))\,.
$$
Now, this is exactly the respond of $h^2$ to a source with support
along the curve $C$, and constant value $1/2\pi$ i.e.
\[
\rho(\vec x)={1\over 2\pi}\int_C{\delta(\vec{x}-
\vec{c}(\gamma))}\, .
\]
In fact, it can be checked that the above density produces the well
known examples of $AdS_3\times S^3$ and pp-wave in six dimensions, where 
circular profiles and infinite straight line are used respectively.

Therefore, we have arrived to the final form that completely
defines the source for regular bubbling solutions. Simply replace
the boundary constraint behaviour found before to obtain\footnote{Where $z$
is reduced to a line integral using Stokes theorem.}
\begin{eqnarray}
\label{zh}
z(\vec x,y)  =  \frac{1}{2\pi} \oint_C \frac{\vec{n}(\vec{x}-
\vec{c})}{(\vec{x}-\vec{c})^2+y^2} +\beta \, , \quad h^2(\vec x,y)  =
\frac{1}{2\pi} \oint_C \frac{1}{(\vec x-\vec c)^2+y^2}\,,
\end{eqnarray}
where $\beta$ is related to the boundary behaviour at infinity, 
${\bf C}$ is a general curve dividing the plane expanded by
$(x^1,x^2)$ into two regions and $\vec n$ is a unit normal vector
pointing to the region where $z=-1/2$. Notice that the integrals are
re-parametrization invariant, as we should expect for a
geometrical solution in gravity.

Hence, we have seen how the two different initial functions
appearing on the ansatz, become related via the same boundary
condition, that comes in terms of a closed curve. Here the
bubbling is realized through changing the shape of the curve. Notice
that there is no reason to have a single connected curve, and
that in general we will have disconnected closed curves as sources
for $h^2$ and $z$.

Next, we compute the flux $f$ of the 3-form $H_{(3)}$ on the above
solutions. Basically, we choose the three dimensional hypersurface
 as follows: define a two dimensional surface $\Sigma_2$ on $(y,x^1,x^2)$
 such that at $y=0$ ends on a closed non-intersecting curve $\Sigma_1$
 that encloses the curve ${\bf C}$, defining a disc $D_2$ containing ${\bf C}$.
 Then, define $\Sigma_3$ as the fibration
 of $\Sigma _2\times S^1$, where $S^1$ is the circle not contracting to
 zero size on $\Sigma_1$ (see figure 1). Computation of $f$ gives
\begin{eqnarray}
f&=& \int_{\Sigma} H_{(3)}  =  - \pi \int_{\Sigma_2}
\tilde{F}_{(2)} =  2 \pi \int_{\Sigma_2} *_3 y d(h^2)\nonumber \\
 &= & 4 \pi^2 \int_{D_2} {\rho (x_1,x_2) dx_1 dx_2} \nonumber \\
& = & 2 \pi\, \mathcal{L}\,. \nonumber
\end{eqnarray}
where $\mathcal{L}$ is the length of the curve ${\bf C}$.
\begin{figure}
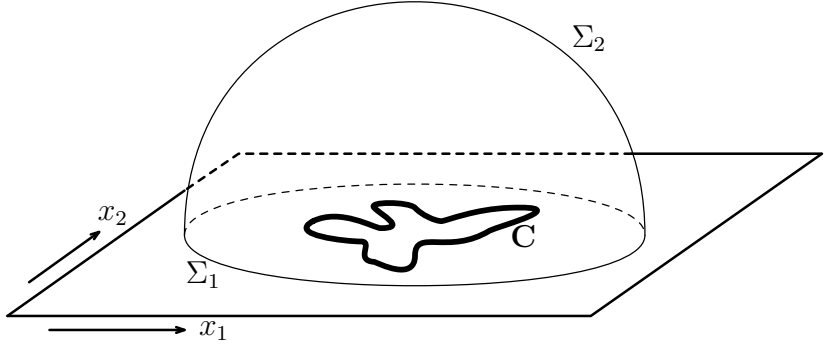
 
\label{fluxfig}
\begin{center}
\begin{texdraw}
  \drawdim cm \setunitscale 1.8 \linewd .02 
  \move(-2.01 -.1) \lvec(2.3 -.1) \lvec(4.01 1.1) \lvec(2.6 1.1) \lpatt(.05 .05) \lvec(-.3 1.1) \lvec(-.7 .83) 
  \lpatt() \lvec(-2.01 -.1)
  \move(-1.85 .15) \arrowheadsize l:.06 w:.04 \arrowheadtype t:V \avec(-1.33 .52) \htext(-1.35 .57){$x_2$}
  \move(-1.7 -.2) \arrowheadsize l:.06 w:.04 \arrowheadtype t:V \avec(-.7 -.2) \htext(-.6 -.28){$x_1$}
  \linewd .01
  \move(-.7 .5) \lpatt(.05 .05) \clvec(-.7 1)(2.7 1)(2.7 .5) \lpatt()
  \move(-.7 .5) \clvec(-.7 0)(2.7 0)(2.7 .5)
  \move(-.7 .5) \clvec(-.7 2.8)(2.7 2.8)(2.7 .5)
  \htext(2.15 1.85){$\Sigma_2$}
  \htext(-.7 .1){$\Sigma_1$}
  \linewd .02
  \linewd .04
  \move(.25 .6) 
  \clvec(.45 .7)(.7 .55)(.75 .55)
  \clvec(.95 .55)(.75 .65)(.7 .7)
  \clvec(.65 .75)(.95 .75)(1 .7)
  \clvec(1.05 .65)(1.15 .6)(1.2 .6)
  \clvec(1.25 .61)(1.45 .69)(1.7 .7)
  \clvec(2.2 .72)(1.63 .57)(1.55 .53) 
  \clvec(1.25 .35)(1 .55)(1 .35)
  \clvec(1 .15)(.73 .29)(.7 .3)
  \clvec(.55 .31)(.7 .45)(.6 .45)
  \clvec(.5 .45)(.05 .5)(.25 .6)
  \htext(1.7 .4){{\bf C}}
\end{texdraw}
\end{center}
\caption{The flux of $H_{(3)}$ is calculated using the 
three dimensional hypersurface $\Sigma_3$, constructed by fibering 
the correspondingly not contracted $S^1$ with the surface $\Sigma_2$.}
\end{figure}

Let us now summarize what we have found up to now. First, the
bubbling ansatz comes in terms of two independent functions
$(h^2,z)$, sourced by independent charge distributions on the plane
expanded by $(x^1,x^2)$. Secondly, once we require regularity, the sources
get identified, in terms of single distributions on a closed curve
${\bf C}$. The total flux $f$ of the above solutions is
proportional to the length of the curve. Based on the above,

\vspace{.5cm}

{\it we define bubbling on $AdS_3\times S^3$ as all the above
regular solutions, constrained to have the same flux, i.e. the same
length but with arbitrary number of disconnected parts of any shape.}

\begin{figure}[h]
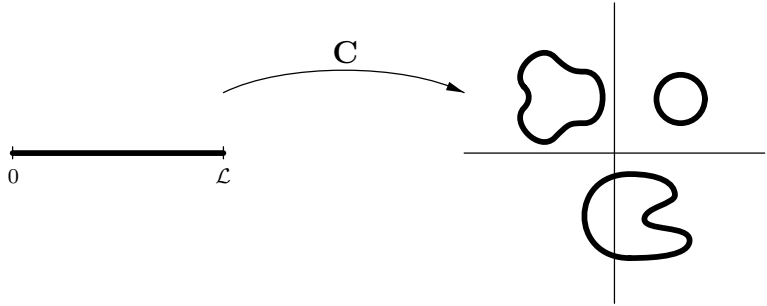

\label{generic}
\begin{center}
  \begin{texdraw} 
    \drawdim cm \setunitscale 4 \linewd .02
    \move(-1.5 .5) \lvec(-.8 .5)
 \linewd .005
    \move(-1.5 .48) \lvec(-1.5 .52)
    \move(-.8 .48) \lvec(-.8 .52)
    \htext(-1.52 .4){\scriptsize{$0$}}
    \htext(-.83 .4){\scriptsize{$\mathcal{L}$}}
    \move(-.8 .7) \clvec(-.6 .8)(-.2 .8)(0 .7)
    \move(-.01 .705) \arrowheadtype t:F \arrowheadsize l:.06 w:.03 \avec(0 .7)
    \htext(-.44 .8){{\bf C}}
    \move(0 .5) \lvec(1 .5)
    \move(.5 0) \lvec(.5 1)
    \linewd .02
    \move(.2 .72)
    \clvec(.15 .77)(.25 .87)(.3 .82)
    \clvec(.35 .77)(.37 .77)(.4 .77)
    \clvec(.48 .77)(.48 .6)(.4 .6)
    \clvec(.35 .6)(.35 .6)(.3 .55)
    \clvec(.25 .5)(.15 .6)(.2 .65)
    \clvec(.22 .67)(.22 .7)(.2 .72)
    \move(.55 .15)
    \clvec(.35 .15)(.35 .43)(.55 .43)
    \clvec(.65 .43)(.7 .4)(.7 .36)
    \clvec(.7 .33)(.6 .32)(.6 .28)
    \clvec(.6 .24)(.75 .26)(.75 .21)
    \clvec(.75 .16)(.65 .15)(.55 .15)
    \move(.72 .68)
    \lcir r:.08
  \end{texdraw}
  \end{center}
  \caption{Generic bubbling profile: fixed the length $\mathcal{L}$, 
we can draw any set of disconnected closed curves.}
\end{figure}

Notice that the above construction is similar, but different to the
bubbling on $AdS_5\times S^5$, where fixing the flux was equivalent
to fix the area of the drop. Here, is the length what matters!


\section{D1/D5 inputs into bubbling}
\label{d1d5}

In this section, we study the relation between the above bubbling
family of solutions and the well known D1/D5 solutions found by
Mathur et al \cite{Lunin:2002bj}. The idea is to set the dictionary to the
dual CFT theory. To make the comparison simpler it is
convenient to rewrite the metric (\ref{bubbling1}) in terms of new
angular variables $(\alpha,\phi)$ as
\begin{eqnarray}
\label{bubbling2} &&ds_6^2  =   -h^{-2} \left[ \left(dt + V
\right)^2 - \left(d\alpha + B \right)^2\right] + h^2 \left( dy^2 +
y^2d\phi^2 + \delta_{ij}\,dx^i dx^j\right)\,,\nonumber \\
&&dB=-*_4dV \,,\quad B=zd\phi\,, \nonumber \\
&&\alpha={1\over 2}(\theta_1+\theta_2)\,,\quad
\phi=(\theta_1-\theta_2)\,,
\end{eqnarray}
where, in this section, we only show the metric for brevity, and
have eliminated $(G,\chi)$ in terms of $(h^2,z)$. Also, $*_4$ is
the four dimensional Hodge dual, acting on $\{y,\phi,x^1,x^2\}$.

Now we turn our attention to the general near horizon metrics of
the D1/D5 system (see for example \cite{Lunin:2002iz})
\bea
\label{md1d5}
ds^2 &=& (f_1
f_5)^{-{1\over 2}} \left[
-(dt-A_Idx^I)^2+(d\alpha+B_Idx^I)^2\right] - (f_1 f_5)^{1\over 2}
(\delta_{IJ}dx^I dx^J)\,,\nonumber \\
dB & = & - *_4 dA\,, \label{dBdC}\nonumber \\
e^{2\Phi} & = & f_1 f_5^{-1}
\eea
with Hodge dual $*_4$ acting on $I=1,2,3,4$ and 
$$
f_5 ={Q_5\over l}\int_{0}^{l} {dv\over
 |{\vec x}-{\vec F}|^2}~,
~~ f_1 ={Q_5\over l}\int_{0}^{l} {|\dot{\vec F}|^2\, dv\over
 |{\vec x}-{\vec F}|^2}~, \quad
A_i = -{Q_5\over l}\,\int_{0}^{l} {\dot{\vec F}_i\, dv\over
 |{\vec x}-{\vec F}|^2}~.\label{profiles}
$$
${\vec F}$ is a vector field that describes the embedding of the
closed curve ${\bf F}$ along the $x^I$ space directions, ${\dot{\vec F}}$
is the derivative of ${\vec F}$ respect to the parameter $v$.
In this parametrization we have that $v=(0,l)$ and
$l=2\pi R' Q_5$ where $Q_5$ is the D5-brane charge, $R'$ is the radius
of the U-dual compact $S^1$ (that here is redefined to be 1), with angular 
variable $\alpha$. 
Notice that these solutions are not invariant if we change the 
parametrization of the curve ${\bf F}$, since there is physical content on the
above. The D1-brane charge $Q_1$ and the angular momentum 
$J_{IJ}$ are given by\footnote{The angular momentum is computed in the full 
solution, before the near horizon limit is considered.}
$$
Q_1={Q_5\over l}\int_{0}^{l} {|\dot{\vec F}|^2\,
dv} ~, ~~J_{IJ}={Q_5\over l}\,\int_{0}^{l} { ( F_I \dot F_J-
F_J\dot F_I)\, dv}\,.
$$

The above solutions correspond to semiclassical configurations of
the D1/D5 system where ${\bf F}$ is related to the profile of the
U-dual bound state of $Q_5$ fundamental strings and $Q_1$
units of momentum. In \cite{Lunin:2002iz} regularity conditions 
for this family of solutions were studied, finding that 
$|\dot{ \vec F}|$ has to be different from zero, and ${\bf F}$ not 
self-intersecting. Therefore, to make contact with the bubbling 
solutions of (\ref{bubbling2}), we restrict them as follows:
\begin{itemize}
\item $|\dot {\vec F}| =\,constant \neq 0$ ,
\item ${\vec F}$ is constrained to live in two dimensions only,
\item $\{x^I\}=\{y, \phi ,x^2,x^1\}$ ,
\item $h^2=(f_1f_5)^{1/2}$ , $V_I=-A_I$ ,
\item we identify the curves ${\bf C}$ and ${\bf F}$ on a fixed 
parametrization.
\end{itemize}

The first point is necessary to work with minimal supergravity
and regular solutions, the second reduces the solutions to
the case where $j =\bar j$, and the curve ${\bf F}$ to be embedded
into a plane. The third and fourth are just the obvious
identifications of coordinates and functions, while the last identifies
the curves. It is important to not only identify the curves, but to
fix a parametrization once and for all, due to the fact that \ref{md1d5}
depends explicitly on this parameter.

\vspace{1cm}\noindent \textsc{\bf AdS/CFT dictionary}
\vspace{.5cm}\\
At this point, we are ready to derive the
dictionary between gravity solutions and CFT states. Let us
begin with some basic facts and definitions. First of all, since
$|\dot{\vec F}|$ is constant, we get that $|\dot{ \vec
F}|=\sqrt{Q_1/Q_5}$. Then, we compute the total flux to obtain
that $f=4\pi^2\sqrt{Q_1Q_5}$. Therefore, we learn that the
parametrization of ${\bf C}$ should be identical to the parametrization
of ${\bf F}$. Hence, in gravity, the rapidity we circulate on the
curve fixes the density of D1-branes, while the length of the
total circulation fixes the product of the number of D1 and D5
branes.
In other words, once we have set the parametrization in the bubbling ansatz,
we have fixed the number of D1 and D5 branes in the system.

The supergravity chiral primaries were studied in 
\cite{Maldacena:1998bw,deBoer:1998ip,Deger:1998nm}. Among
them, there is a special family with $j=\bar j=1,3/2,\cdots$, that
produces fluctuations on the metric of the $3$-sphere,
associated with the anti-self dual part of the 3-form $H_{(3)}$. In
\cite{Lunin:2002iz} it was noticed that such states are related to changes on
the shape of the profile ${\bf F}$. Since this is the only freedom
left in our supergravity solutions, we conclude that these are
precisely the chiral primaries that we can probe or excite within the
bubbling framework.

Following Mathur et al \cite{Lunin:2001pw}, we use the twist operators
$(\sigma_n^{++},\sigma_n^{+-},\sigma_n^{--},\sigma_n^{-+})$\footnote{These 
operators expand a $(1/2,1/2)$ representation of
$SU(2)\times SU(2)$. For more details see the original paper.} to
describe the chiral primaries under study. Here $\sigma_n$
permutes cyclically the ends of each D-string in the
system. Since we have a total $N=Q_1Q_5$ of such strings, $n$ runs
from $1$ to $N$. Due to the fact that we consider the
case $j=\bar j$, we are allowed to use only
$(\sigma_n^{++},\sigma_n^{--})$. Also, $\sigma_1^{--}$ has the lower
conformal dimension, therefore the vacuum configuration in the
twist sector is given by the operator $[\sigma_1^{--}]^N$, while a
general configuration is given by the product of
$[\sigma_{n_i}^{--}]^{m_i}$ and $[\sigma_{n_j}^{++}]^{m_j}$
subject to the constraint $\sum (n_im_i +n_jm_j)=N$.

\vspace{.8cm}
Let us consider the vacuum configuration. First we recall that
the vacuum maximizes the total spin (since we have $N$ aligned spin
$1/2$ short strings). From the bubbling point of view, the vacuum
also maximizes angular momentum or, better, corresponds to the
curve of fixed length that maximizes its angular momentum: the
circle. In detail, we find the vacuum by considering first a
circular profile
\bea
&&\vec F= a \cos(wv) \hat e_1 + a \sin(wv)
\hat e_2 \cr &&|\dot {\vec F}|=aw \nonumber
\eea
where $(\hat e_1,\hat e_2)$
are the unit vectors on the two dimensional plane defined by
$(x^1,x^2)$. Then, using the two constraints
$f=4\pi^2\sqrt{Q_1Q_5}$, and $|\dot{ \vec F}|=\sqrt{Q_1/Q_5}$, we
obtain that $a=\sqrt{Q_1Q_5}$ and $w=2\pi/l$. Inserting the resulting
curve into equation (\ref{zh}), we get
$$
h^2={a \over
{\sqrt{(y^2+r^2+a^2)^2-4a^2r^2}}}\quad,\quad z={1\over
2}{y^2+r^2-a^2 \over {\sqrt{(y^2+r^2+a^2)^2-4a^2r^2}}}
$$
where
$\delta_{ij}dx^idx^j=dr^2+r^2d\psi$. Using the following change of
coordinates $y=a\,\sigma\sin \theta $,
$r=a\sqrt{\sigma^2+1}\,\cos \theta $, we recover $AdS_3\times S^3$
metric in global coordinates
\bea
ds^2 & = & L^2[-(1+\sigma^2)dt^2+{1\over
{1+\sigma^2}}d\sigma^2+\sigma^2d\theta_1^2\,] \nonumber \\ \nonumber \\
&& + L^2\left( d\theta^2+\cos^2\theta d\psi^2+\sin^2\theta d\theta_2^2 \right), \nonumber
\eea
with $L = (Q_1Q_5)^{1/4}$.

\vspace{.8cm}
The other celebrated example is the pp-wave solution that, as in
the case of bubbling for D3-branes, can be obtained by focusing on
local part of the curve ${\bf C}$, which looks like a straight
infinite line. In this limit, we get
$$
h^2={1 \over{2\sqrt{(y^2+x_2^2)}}}\quad,\quad z={x_2 \over {2\sqrt{(y^2+x_2^2)}}}\,,
$$
where $(x_1,x_2)$ are adapted coordinates, such that they are perpendicular and
parallel respectively to the curve ${\bf C}$. Then, changing coordinates 
like in (\ref{pp-wave}), we obtain the familiar metric
$$
ds^2 = -(r_1^2+r_2^2)dt^2-2dtdx_1 +dr^2_1+dr^2_2+r_1^2d\theta_2+r_2^2d\theta_1\,.
$$

\vspace{.8cm}
Once we have set the dictionary for the vacuum, we can start to
study excited states on this sector of the theory. We point out
that we can very well consider disconnected curves like those of
figure (3), remaining on minimal supergravity domains. Hence, a
little puzzle comes into our minds: these configurations seem to
be related to giant gravitons, but the latter were associated to
different supergravity solutions with non-trivial
dilaton behaviour in \cite{Lunin:2002bj}, and therefore should be out of the
minimal supergravity theory.
To solve this puzzle more studies on the disconnected curves
need to be done, that are left for future works.

\begin{figure}[t]
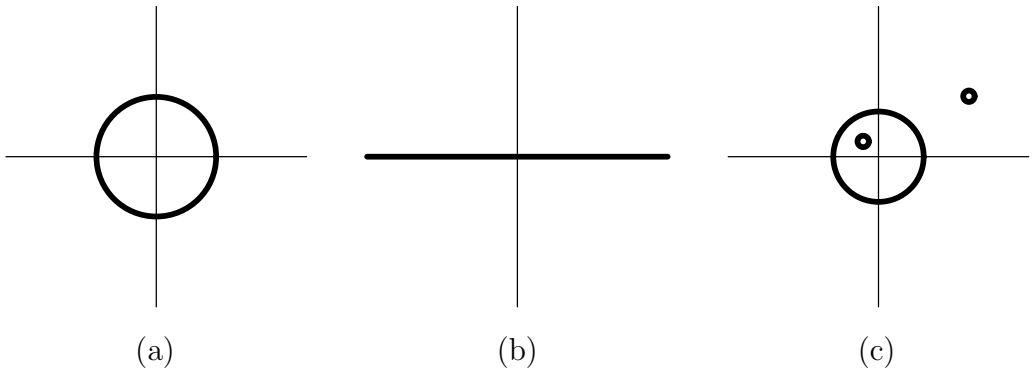
                                                    
  \begin{texdraw}
    \drawdim cm  \setunitscale 4 \linewd .005
    \move(0 .5) \lvec(1 .5)
    \move(.5 0) \lvec(.5 1)
    \move(.5 .5) \linewd .02 \lcir r:.2 
    \htext(.43 -.2){(a)}
    \move(1.2 .5) \lvec(2.2 .5)
    \linewd .005 \move(1.7 0) \lvec(1.7 1)
    \htext(1.63 -.2){(b)}
    \move(2.4 .5) \lvec(3.4 .5)
    \move(2.9 0) \lvec(2.9 1)
    \move(2.9 .5) \linewd .02 \lcir r:.15
    \move(2.85 .55) \lcir r:.02
    \move(3.2 .7) \lcir r:.02
    \htext(2.83 -.2){(c)}
  \end{texdraw}
  \caption{three possible configurations of bubbling: (a) the vacuum 
geometry $AdS_3\times S^3$, (b) its Penrose limit, corresponding 
to the pp-wave solution, and (c) a superposition of disconnected 
closed curves in and out of the main big circle, that may be related to
giant graviton configurations in $S^3$ and $AdS_3$ respectively. 
In fact, this is the type of figures we get if we write giant graviton 
test branes in bubbling coordinates, replacing the small circles by points.}
\end{figure}


\section{Bubbling and non-regular solutions}
\label{cft}

In previous sections, we set the rules for bubbling in the D1/D5
system, giving also the dual CFT chiral operators that can be
probed. We could now research on physical implications of this
sector of the theory. Nevertheless, we decided to postpone such a
study for better times, and to work out, instead, the appearance (and the
CFT description, if any) of non-regular solutions. Therefore, in
this section we will be working with a slight generalization of the
bubbling ansatz, relaxing the regularity conditions. In this form,
we are able to recover other sectors on the CFT theory (not included
into the regular bubbling ansatz that nevertheless is interesting in its
own rights) containing for example conical defect metrics.

From the D1/D5 brane perspective, among the semiclassical
solutions, there is some room for generalization: we can 
relax the constraints on ${\bf F}$, by allowing either $\dot{\vec F}=0$
along the curve or self intersections. Also we can consider
superposition of regular profiles. On the other hand, from the
perspective of bubbling, we have a lot of room to play with.
Basically, as soon as we give up the regularity conditions, we have
two independent sources. Here, we will conserve the notion of {\em lines of
charge as sources}, and work with asymptotically $AdS_3\times S^3$
solutions and with round profiles such that we gain an extra Killing
direction that simplifies the studies\footnote{This is a sort of minimal
generalization, but obviously other more general deviations from the 
regular conditions could be considered. Here we are just probing 
a reduced region of the space of possible solutions.}. 
We have chosen to parameterize the four possible cases in terms 
of $2\pi\rho_0$ and $\Delta_- z$, where $\rho_0$ is the source density 
for $h^2$ and $\Delta_\pm z=z_{I}\pm z_{II}$ ($z_{I}=1/2$ from 
the asymptotic boundary conditions). Then we have 
\bea 
\label{cases} 1)
&&\Delta_- z =2\pi\rho_0 \neq 1 \,, \cr 2) &&\Delta_- z =1
\,\hbox{and}\;2\pi\rho_0 \neq 1 \,,\cr 3) &&\Delta_- z \neq
1\,\hbox{and}\;2\pi\rho_0 =1\,, \cr 4) &&\Delta_- z \neq
1\,,\,2\pi\rho_0 \neq 1\,\hbox{and}\; \Delta z \neq 2\pi\rho_0\,.
\eea 
while the regular case is recovered by setting $\Delta_- z = 2\pi\rho_0 =1$.

The fact that we are considering only circular profiles of radius
$a$ reduces the form of $z,h^2$ and $V_\psi$ ( the additional 
Killing vector sets $V_r=0$) to 
\bea 
\label{zh2}
&&z(r,y)={1\over 2} \left[ \Delta_+z + {\Delta_-z
(y^2+r^2-a^2)\over \sqrt{y^2+r^2+a^2)^2-4r^2a^2}}\right] \,,\cr
&&h^2(r,y)= {L^2\over  \sqrt{y^2+r^2+a^2)^2-4r^2a^2}} \,, \cr
&&V_\psi  =  \frac{1}{2} \Delta_- z \left[ 1 -
\frac{R^2+y^2+a^2}{\sqrt{(R^2+y^2+a^2)^2-4 R^2 a^2}} \right]\,, \nonumber
\eea 
where $L^2=\sqrt{Q_1Q_5}$ . Since we work with a total fixed
flux $f=4\pi^2\sqrt{Q_1Q_5}$, the radius of the circular profile is
constrained to $a={L^2 /2\pi \rho_0}$. We also change to
$AdS$-adapted coordinates 
$$
y=L^2\sigma \sin\theta\quad ,\quad
r=l^2\sqrt{\sigma^2+\alpha}\cos\theta 
$$ 
where $\alpha=1/(2\pi\rho_0)^2$, obtaining
\bea
\label{zh3}
&&z(\sigma,\theta)={1\over 2} \left[ \Delta_+z + \Delta_-z
{(\sigma^2-\alpha\sin^2\theta) \over
(\sigma^2+\alpha\sin^2\theta)}\right]\,, \cr 
&&h^2(\sigma,\theta)=
{1\over  L^2(\sigma^2+\alpha\sin^2\theta))}\,,\cr\cr 
&&V_\phi =  -
\frac{\alpha\Delta_- z\cos^2 \theta}{(\sigma^2 +\alpha\sin^2
\theta)}
\eea
and since $z_I=1/2$, only one of the two $\Delta_\pm
z$ is independent, rendering the whole supergravity solution a
function of only $\Delta_- z$ and $\alpha$.

We would like to identify which of the above supergravity
solutions are {\em physical solutions}, that is to say which ones
have a CFT dual configuration. At first sight, it seems that we
are out of the D1/D5 system: after all, there should be only one
source in the physical situation (here, in general, we are dealing 
with two independent sources). Nevertheless, we know that there
are, within the bubbling ansatz, non-regular solutions that
correspond to condensates of otherwise regular solutions with CFT
dual states\footnote{For example, in the D3-brane case we have the
superstar solution \cite{Cvetic:1999xp}, that can be
understood as a sort of average over the metrics produced by 
the giant graviton distribution \cite{Myers:2001aq}. In this case, we relax
the two possible boundary values of $z$ associated to white and
black in the figures, such that we also have grey tonalities,
associated to regions where the average is considered and therefore
the source is effectively smeared.}.

In the D1/D5 system we could change the value of $\Delta_-z$
trying to see this as the result of an {\em average} over delocalized
sources. The above procedure could also be implemented on the
boundary, changing the local amount of D1-branes we locate on this
curve, that therefore varies the value of $2\pi\rho_0$. Hence, we
just can not rule out the family of solutions (\ref{zh3}) without
further studies. Nevertheless, certainly there are ranges of the
parameters $\Delta_-z$ and $2\pi\rho_0$ that do not have any associated
CFT dual configuration. Hopefully, the above non-physical solutions should 
be associated to CTC or other types of pathological behaviours.

\begin{center}
\begin{figure}
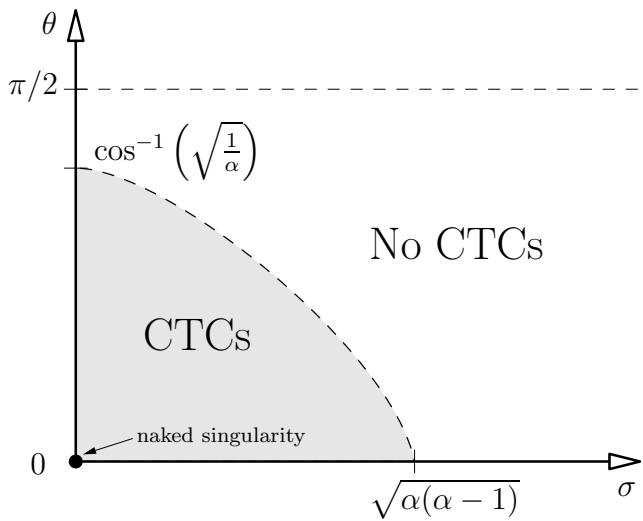

\label{ctcfig}
\begin{center}
  \begin{texdraw} 
    \drawdim cm \setunitscale 3 \linewd .01
    
    \linewd .005
    \lpatt(.05 .05)
    \move(.5 .5) \lvec(.5 1.8)
    \clvec(.9 1.8)(1.9 1)(2 .5)
    \lvec(.5 .5 )
    \lfill f:.9
    \lpatt()
    \linewd .01
    \move(.5 .5) \avec(.5 2.5)
    \move(.5 .5) \avec(3 .5)

    \linewd .001
    \lpatt(.05 .05)
    \move(.5 2.15) \lvec(3 2.15) 
    \lpatt()
    \move(.5 .5) \fcir f:0 r:.03

     \htext(0.2 2.1){$\pi/2$}
     \move(.45 2.15) \lvec(.55 2.15)
     \htext(.58 1.75){$\cos^{-1}\left(\sqrt{1\over \alpha}\right)$}
     \move(.45 1.8) \lvec(.55 1.8)
     \htext(0.35 2.4){$\theta$}

     \htext(1.8 0.25){$\sqrt{\alpha(\alpha-1)}$}
     \move(2 .45) \lvec(2 .55)
     \htext(.8 1){\Large CTCs}
     \htext(1.8 1.4){\Large No CTCs}
     \htext(0.3 .45){0}
     \htext(2.9 .35){$\sigma$}
     \move(.75 .6) \arrowheadsize l:.06 w:.03 \arrowheadtype t:F \avec(.53 .53)
     \htext(.77 .57){\scriptsize{naked singularity}}
  \end{texdraw}
  \end{center}
  \caption{The picture shows the CTC region for fixed $\alpha$ in the $(\theta,\sigma)$ plane.}
\end{figure}
\end{center}

In fact, one of the general features of the solutions (\ref{zh3}),
is that there are CTC. Take for example the $g_{\psi\psi}$
components of the metric
\bea
g_{\psi \psi} & = & L^2\frac{\cos^2 \theta}{\sigma^2 +\alpha\sin^2
\theta} \left[ \sigma^2 + \alpha \left( 1 -
\alpha(\Delta_-z)^2\cos^2 \theta\right) \right]\nonumber
\eea
from which we can see that there are regions in space-time with
CTC if
$$
\alpha>(\Delta_-z)^{-2}\quad \hbox{or if you prefer}
\quad |\Delta_-z|>|2\pi\rho_0| \,.
$$
These regions are defined by the equation
$$
\sigma^2<\alpha[\alpha(\Delta_-z)^2\cos^2\theta-1]
$$
and the metric has a naked singularity for $\theta=\sigma=0$ (see figure \ref{ctcfig}).

It would be very interesting to translate the above supergravity
range of parameters into CFT parameters, to actually understand if
such pathological solutions are or not in the CFT theory.
Unfortunately, these supergravity solutions would be related to
averages over microscopical configurations, and it is not assured
that we will discover how the average was done, and over which
microstates. Nevertheless, we have been able to find the CFT dual
configuration for the first two cases listed in (\ref{cases}),
respectively corresponding to conical singular metrics and to the
Aichelburg-Sexl type metric found in \cite{Lunin:2002fw}. In both cases, we
found CFT duals only for the non pathological regimes, enforcing a
protection mechanism curing gravity.

\vspace{1cm}
\noindent
\textsc{\bf Case  $\Delta_-z =  2\pi\rho_0\neq 1$}
\vspace{.5cm}\\
Let us consider the first case, i.e. where $\Delta_- z=2\pi
\rho_0\neq 1$. We have chosen to parametrize it as follows,
$$
\alpha=(1+Q)^2\quad,\quad \Delta_-z={1\over(1+Q)}\,,
$$
where, in general, Q runs from $-\infty$ to $\infty$. The resulting form
of the metric is
\bea
ds^2 & = & L^2 [-((1+Q)^2+\sigma^2)dt^2 +
{d\sigma^2\over((1+Q)^2+\sigma^2)}+ \sigma^2d{\bar \theta}_1^{\,2} \nonumber \\ \nonumber \\
&& \quad\quad +\, d\theta^2 +\cos^2\theta d{\bar \psi}^2 + \sin^2\theta d{\bar \theta}_2^{\,2}
] \quad\quad \nonumber
\eea 
where
$$\bar \theta_1=\theta_1 \quad, \quad \bar \theta_2=\theta_2
+Q\theta_1\quad ,\quad\bar \psi=t+\psi$$
This metric corresponds to a local $AdS_3\times S^3$ space-time,
with a conical singularity. If $0 < (1+Q)< 1$ we have a conical
defect, but if $1<(1+Q)<2$ we have a conical excess\footnote{Other 
values of $Q$ are irrelevant, due to the
periodicity of $\bar \psi$.}.

To make contact with the D1/D5 system, we have to set the correct
parametrization of the curve ${\bf C}$. We found that the form of
the curve and the flux constraint define uniquely the
parametrization to be given by 
\bea 
&\vec F= a \cos(w_Qv) \hat e_1 +
a \sin(w_Qv)\hat e_2 \quad,\quad v=(0,l)&\cr \cr
&a=\sqrt{Q_1Q_5}(1+Q)\quad,\quad w_Q={2\pi \over l(1+Q)}& \nonumber 
\eea 
At this point, we have to introduce the physical constraint that the
curve ${\bf F}$ is actually closed. This is only achieved if
$Q=(1-m)/m$ with $m \in {\mathbb N}_0$. In other words, we can
find a corresponding configuration on the CFT side, as long as $Q
\in (-1,0)$. These allowed cases produce only conical defect
metrics, with deficit angle $\delta{\bar\theta}_1=2\pi Q$. Of
course this case is not regular since we have self intersections
on {\bf F}.

The relation between conical defects and CFT operators has been
studied before for the D1/D5 system \cite{Lunin:2002fw}. 
The dual associated operator is given by 
$$
[\sigma_m^{--}]^{N/m} 
$$ 
and produces a deficit angle $\delta{\bar\theta}_1=2\pi(1-1/m)$. 
Notice that the system is excited to higher energy state that in 
the U-dual P/F1 picture corresponds to the harmonic $w_m = 2\pi m/l$.

\begin{figure}[t]
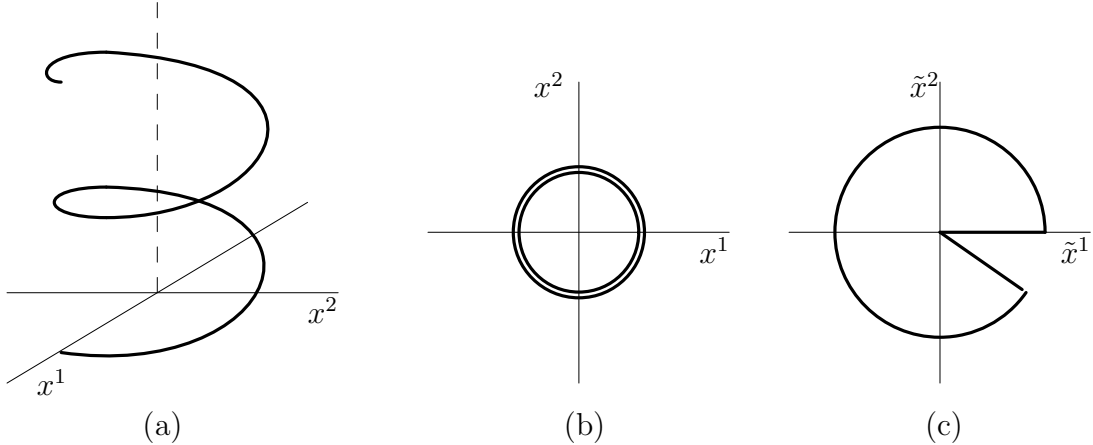

 \begin{texdraw}
  \drawdim cm  \setunitscale 4 \linewd .001
 \move(0 .5) \lvec(1.1 .5)
 \move(0 .2) \lvec(1 .8)
 \lpatt(.05 .05)
 \move(.5 .5) \lvec(.5 1.5)
 \lpatt()
 \linewd .01
 \move(0.18 0.3) \clvec(0.9 0.2)(1.18 .83)(.33 .85)
 \move(0.33 .75) \clvec(0.9 0.75)(1.18 1.25)(.33 1.3)
 \move(0.33 .75) \clvec(0.1 .75)(.1 .85)(.33 .85)
 \move(0.18 1.2) \clvec(0.1 1.2)(.1 1.3)(.33 1.3)
 \htext(.1 .16){$x^1$}
 \htext(1 .4){$x^2$}

 \linewd .001
 \move(1.4 .7) \lvec(2.4 .7)
 \move(1.9 .2) \lvec(1.9 1.2)
 \move(1.9 .7) \linewd .01 \lcir r:.2
 \lcir r:.217
 \htext(1.75 1.15){$x^2$}
 \htext(2.3 .6){$x^1$}

 \linewd .001
 \move(2.6 .7) \lvec(3.6 .7)
 \move(3.1 .2) \lvec(3.1 1.2)
 \move(3.1 .7) \linewd .01 \larc r:.35 sd:0 ed:325
 \move(3.1 .7) \lvec(3.45 .7)
 \move(3.1 .7) \lvec(3.375 .51)
 \htext(3 1.15){${\tilde x}^2$}
 \htext(3.5 .6){${\tilde x}^1$}

\htext(.45 0){(a)}
\htext(1.85 0){(b)}
\htext(3.05 0){(c)}
 \end{texdraw}

\caption{ (a) shows a profile that circulates more than once with
frequency $w_Q$. (b) shows its projection into the $(x^1,x^2)$-plane,
 where the double line is stressing the self interaction of the curve.
 (c) shows the same profile in terms of new variables 
${\tilde x}^i={1\over(1+Q)}x^i$. In this new frame, the radius 
of the circle is $L^2$ as in the vacumm case but the curve fails 
to close, with deficit angle $\delta \bar{\theta}_1=2\pi Q$.}
\end{figure}

So, we have seen how to recover conical defect metrics within the
bubbling picture, in a completely independent way, compared to how such metrics
were found in the D1/D5 system. Also, we have gained more physical
input into other related supergravity solutions, the conical
excess. From the above considerations, these solutions are not in
the spectrum of the dual CFT theory and, therefore, should be ruled
out as un-physical solutions\footnote{One way to observe the
impossibility of such states is by noticing that they imply the
existence of harmonics of lower energy than $w_1=2\pi/l$, which is
the ground energy level, and corresponds to an initial working
assumption. All the above is in the U-dual P/F1 system.}. This is a
typical behaviour of string theory, which teaches us that not all
the supergravity solutions are to be labeled as physical (keeping 
in mind that  gravity is just an effective theory). 
In this particular example, the result is not so unexpected, since conical 
excess can be thought as the response of space-time to negative 
energy point particles, that even at the classical level are somehow 
related to pathologic behaviours. Nevertheless, this example shows the
power of the AdS/CFT duality since it leaves no room for doubts
about the physical meaning of these metrics.

\vspace{.5cm} \noindent
 \textsc{\bf Case $\Delta_-z=1\ , \ 2\pi\rho_0 \neq 1$}
\vspace{.5cm}\\
In this second case we set the parametrization as follows
$$
\alpha=(1+Q) \quad,\quad \Delta_-z=1\,, 
$$  
where $Q$ runs from $-1$ to $\infty$. The resulting metric is
\begin{eqnarray}
\label{as}
ds^2 & = & - L^2 (\sigma^2 + \alpha) dt^2 + \frac{L^2}{\sigma^2 
+ \alpha} d\sigma^2 + L^2 \sigma^2 \frac{\sigma^2 
+ \alpha \sin^2 \theta}{\sigma^2 + \alpha^2 \sin^2 \theta} 
d\theta_1^2 \nonumber \\
\nonumber \\
&   & + L^2 d\theta^2 + L^2 \sin^2 \theta \frac{\sigma^2 
+ \alpha^2 \sin^2 \theta}{\sigma^2 + \alpha \sin^2 \theta} 
\left( d\theta_2^2 - \frac{(1-\alpha)\sigma^2}{\sigma^2 + 
\alpha^2 \sin^2 \theta} d\theta_1^2 \right)^2 \nonumber \\
\nonumber \\
&   & + L^2 \alpha \cos^2 \theta d\tilde{\phi}^2 + 
\frac{L^2 \cos^2 \theta}{\sigma^2 + \alpha \sin^2 \theta} 
(1-\alpha) (\sigma^2 + \alpha^2) (d\tilde{\phi} - dt)^2 
\end{eqnarray}
that corresponds to the Aichelburg-Sexl metric found in
\cite{Lunin:2002fw}. This can be easily seen by comparing the form of
$(h^2,V_\psi)$ with the corresponding metric functions given in
\cite{Lunin:2002bj} (see equations (3.18) (3.20) and use $q=-Q$). 
These solutions were found in the D1/D5 framework by smearing over a
single turn on the circular profile ${\bf F}$ a large number of
{\em bits of the curve} that remain at a fixed point in space-time,
while we move on the curve of parameter $v$. Therefore we are
changing the density of D1-branes in the resulting average curve (see
\cite{Lunin:2002bj} for a complete construction of this solution).

The above metrics are conjectured to be dual to operators of the
form
$$
[\sigma_{1}^{--}]^p\sigma_{n_1}^{--}\sigma_{n_2}^{--}\ldots\sigma_{n_k}^{--}
$$
with $n_i\gg 1$. The $n_i$'s correspond to {\em bits} on the
U-dual F-string profile, that remain constant at a fixed point in
space-time, while we move on the parameter $v$ along the curve.
The average on the position corresponds to a distribution with a
large dispersion on $n_i$. Then, large numbers of such fixed
{\em bits} will translate into a reduction of the radius $a$ of the circle
described by the profile ${\bf F}$, since less {\em bits} will be left 
to close the curve. In the above sense, radii larger 
than the radius of the vacuum configuration are impossible
to be constructed in the CFT theory. Therefore we conclude that the
solutions with $Q\in (-1,0)$ (or if you prefer $\alpha>1$) are not
physical, since there is no dual CFT configuration. At this point,
we stress that precisely when $\alpha>1$, the supergravity
solutions present CTC! Therefore this is another example of
chronology protection implemented by string theory\footnote{Also,
it is interesting to notice that in the asymptotic regions of large
$\sigma$, these metrics (\ref{as}) suffer from conical alike defect
or excess, depending on whether $\alpha$ is less or bigger than
one, reinforcing our previous conclusions.}.

\begin{figure}
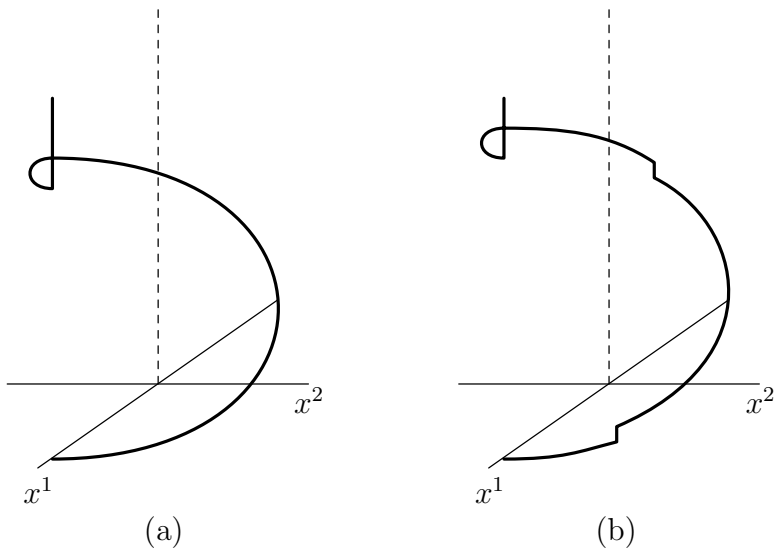
                                            
\begin{center}
  \begin{texdraw}
    \drawdim cm \setunitscale 2 \linewd .02 
    \linewd .01 
    \lpatt(.05 .05) \move(1.7 1.5) \lvec(1.7 4) \lpatt()
    \move(.7 1.5) \lvec(2.7 1.5) 
    \move(.9 .94) \lvec(2.5 2.06)
    \linewd .02
    \htext(.8 .7){$x^1$}
    \htext(2.6 1.3){$x^2$}
    \move(1 1)
    \clvec(3 1)(3 3)(1 3)
    \clvec(.8 3)(.8 2.8)(1 2.8)
    \lvec(1 3.4)
    \htext(1.6 .4){(a)}
    \linewd .01 
    \lpatt(.05 .05) \move(4.7 1.5) \lvec(4.7 4) \lpatt()
    \move(3.7 1.5) \lvec(5.7 1.5) 
    \move(3.9 .94) \lvec(5.5 2.06)
    \linewd .02
    \htext(3.8 .7){$x^1$}
    \htext(5.6 1.3){$x^2$}
    \move(4 1)
    \clvec(4.4 1)(4.5 1.05)(4.75 1.115)
    \lvec(4,75 1.215)
    \clvec(5.8 1.67)(5.6 2.57)(5 2.87)
    \lvec(5 2.97)
    \clvec(4.7 3.17)(4.4 3.2)(4 3.2)
    \clvec(3.8 3.2)(3.8 3)(4 3)
    \lvec(4 3.4)
    \htext(4.6 .4){(b)}
 \end{texdraw}
\end{center}
  \caption{(a) shows a curve that circulates once around the origin, 
with zero velocity in the last portion of the curve. (b) shows a curve that
has many parts whith zero velocity. The Aichelburg-Sexl metric is obtained by
the limiting situation when the straight bits are smeared along the curve.}
\end{figure}


\section{Summary and discussion}
\label{last}

In this article we have studied the relation between bubbling in
$AdS_3$ and the more conventional construction of microstates in
the D1/D5 system. We have learned a few lessons: the first lesson
tells us how both constructions are related once we concentrate on
regular solutions only. In this case the D1/D5 family of
microstates contains the bubbling solutions as a subset. The
connection is possible due to the collapse (in the bubbling
framework) of the different sources into a single line of charge.
This subset is dual to a particular tower of chiral primaries
operators in the CFT with conformal weights $(1,3/2,\ldots)$.
Therefore, we have studied all the geometries sourced by connected
and/or disconnected closed curves with fixed total length. The
second lesson tells us that this is not the full story and that
non-regular solutions have also a role to play in this framework.
This time, the splitting of the sources (characteristic of the
bubbling picture) is understood (from the D1/D5 family of
microscopic states) as the result of an average over semiclassical
configurations, that effectively smears the string source. The third
lesson is that there are solutions within the bubbling ansatz that
have no counterpart in the D1/D5 family of microstates and,
therefore, do not have a CFT dual. These solutions are artifacts
of the low energy supergravity theory and should be discarded as
non-physical. In particular, in all the non-physical solutions we
have found there are pathologies like CTC. Hence string theory
seems to be acting as a chronology protection agency.

It would be very interesting to connect the $AdS/CFT$
picture with the Liouville theory living at the boundary of
$AdS_3$. In fact, it was shown in \cite{Seiberg:1999xz} that the world-volume
theory of a single D1-brane in $AdS_3$ becomes precisely a
Liouville theory once we approach the boundary. The D1-brane
needs to be rotating in the $S_3$ to become a stable BPS state.
Such states are called giant gravitons and are of importance in
the bubbling framework (at least for the D3-brane case). Now, it
is also known that in the Liouville theory there are normalizable
and non-normalizable states (see \cite{Seiberg:1990eb} for a review). 
The normalizable states have a continuous spectrum bounded from below, 
while the non-normalizable states do not present such a gap. 
Therefore, the theory gets organized as a series of sectors labeled by this 
non-normalizable states plus the tower of normalizable states. 
In \cite{Krasnov:2000ia} non-normalizable states were conjectured to 
be dual to conical defect metrics. At this point we would like 
to recall that, from the bubbling point of view, the CFT theory is 
somehow naturally arranged into regular and non-regular
sectors, in such a way that the non-regular sectors act as different vacua,
while the regular sector can be accommodated as a deformation on
each of these vacua. In terms of bubbling figures we have, for example,
a conical defect vacuum, defined by a circular self-interacting curve, 
and a whole tower of operators, produced by small changes on this 
profile, deforming the circle. Each conical defect metric is defined by a
winding number that separates one sector from the other. So we
believe that these similarities signal common structure and that a
deeper study on giant gravitons on the D1/D5 system deserves
attention, since it may provide a bridge between dual CFT theory of
D1/D5 system and the Liouville theory at the boundary of $AdS_3$.


\section*{Acknowledgments}
\small The authors would like to specially thank M. Caldarelli for many suggestions 
and useful discussions during the research. Also we thank D. Klemm and S. Giusto 
for useful comments.

This work was partially supported by INFN, MURST and by the
European Commission RTN program HPRN-CT-2000-00131, in which M.~B.
and P.~J.~S. are associated to the University of Milan.
\normalsize


\end{document}